\documentclass[]{aastex7}

\newcommand\kms{$\mathrm{km\ s^{-1}}$}

\newcommand\alf{Alfv\'en}
\newcommand\alfic{Alfv\'enic}
\newcommand\alfty{Alfv\'enicity}

\newcommand\du{\delta u}
\newcommand\db{\delta b}

\newcommand\ytr{\gamma_{tr}}
\newcommand\ymax{\gamma_{0tr}}
\newcommand\yrat{\gamma_{tr} / \gamma_{0tr}}

\begin{document}

\title{The impact of Alfv\'enic shear flow on magnetic reconnection and turbulence}

\author[0000-0002-8475-8606]{Tamar Ervin}
\affiliation{Department of Physics, University of California, Berkeley, Berkeley, CA 94720-7300, USA}
\affiliation{Space Sciences Laboratory, University of California, Berkeley, CA 94720-7450, USA}
\email{tamarervin@berkeley.edu} 

\author[0000-0001-9202-1340]{Alfred Mallet}
\affiliation{Space Sciences Laboratory, University of California, Berkeley, CA 94720-7450, USA}
\email{alfred.mallet@berkeley.edu}

\author[0000-0002-5619-1577]{Stefan Eriksson}
\affiliation{Laboratory for Atmospheric and Space Physics, University of Colorado, Boulder, CO 80303,USA}
\email{Stefan.Eriksson@lasp.colorado.edu}

\author[0000-0002-5435-3544]{M. Swisdak}
\affiliation{Institute for Research in Electronics and Applied Physics, University of Maryland, College Park, MD 20742, USA}
\email{swisdak@umd.edu}

\author[0000-0001-6835-273X]{James Juno}
\affiliation{Princeton Plasma Physics Laboratory, Princeton, NJ 08543, USA}
\email{jjuno@pppl.gov}

\author[0000-0002-4559-2199]{Orlando M. Romeo}
\affiliation{Space Sciences Laboratory, University of California, Berkeley, CA 94720-7450, USA}
\email{oromeo@berkeley.edu}

\author[0000-0002-6924-9408]{Tai Phan}
\affiliation{Space Sciences Laboratory, University of California, Berkeley, CA 94720-7450, USA}
\email{taiphan@berkeley.edu}

\author[0000-0002-4625-3332]{Trevor A. Bowen}
\affiliation{Space Sciences Laboratory, University of California, Berkeley, CA 94720-7450, USA}
\email{tbowen@berkeley.edu}

\author[0000-0002-0396-0547]{Roberto Livi}
\affiliation{Space Sciences Laboratory, University of California, Berkeley, CA 94720-7450, USA}
\email{rlivi@berkeley.edu}

\author[0000-0002-7287-5098]{Phyllis L. Whittlesey}
\affiliation{Space Sciences Laboratory, University of California, Berkeley, CA 94720-7450, USA}
\email{phyllisw@berkeley.edu}

\author[0000-0001-5030-6030]{Davin E. Larson}
\affiliation{Space Sciences Laboratory, University of California, Berkeley, CA 94720-7450, USA}
\email{davin@berkeley.edu}

\author[0000-0002-1989-3596]{Stuart D. Bale}
\affiliation{Department of Physics, University of California, Berkeley, Berkeley, CA 94720-7300, USA}
\affiliation{Space Sciences Laboratory, University of California, Berkeley, CA 94720-7450, USA}
\email{bale@berkeley.edu} 


\begin{abstract}
Magnetic reconnection is a fundamental and omnipresent energy conversion process in plasma physics. Novel observations of fields and particles from Parker Solar Probe (PSP) have shown the absence of reconnection in a large number of current sheets in the near-Sun solar wind. Using near-Sun observations from PSP Encounters 4 to 11 (Jan 2020 to March 2022), we investigate whether reconnection onset might be suppressed by velocity shear. We compare estimates of the tearing mode growth rate in the presence of shear flow for time periods identified as containing reconnecting current sheets versus non-reconnecting times, finding systematically larger growth rates for reconnection periods. Upon examination of the parameters associated with reconnection onset, we find that 85\% of the reconnection events are embedded in slow, non-Alfv\'enic wind streams. We compare with fast, slow non-Alfv\'enic, and slow Alfv\'enic streams, finding that the growth rate is suppressed in highly Alfvenic fast and slow wind and reconnection is not seen in these wind types, as would be expected from our theoretical expressions. These wind streams have strong Alfv\'enic flow shear, consistent with the idea of reconnection suppression by such flows. This could help explain the frequent absence of reconnection events in the highly Alfv\'enic, near-Sun solar wind observed by PSP. Finally, we find a steepening of both the trace and magnitude magnetic field spectra within reconnection periods in comparison to ambient wind. We tie this to the dynamics of relatively balanced turbulence within these reconnection periods and the potential generation of compressible fluctuations.
\end{abstract}


\section{Introduction} \label{sec:intro}

Magnetic reconnection is a ubiquitous process in plasma environments whereby magnetic energy is converted to thermal energy, heating the plasma. Evidence for magnetic reconnection is found everywhere in astrophysical \citep[e.g.][]{Hesse-2020} and laboratory plasmas \citep{Ji-2023}. It is a multi-scale process that occurs in varied environments: from the pristine solar wind \citep{Gosling-2005} and large stellar eruptions such as coronal mass ejections and flares \citep{Klimchuk-2006}, to driving geophysical flows and the aurorae in planetary atmospheres \citep{Dungey-1953, Hoyle-1949, Dungey-1961, Jia-2012, Paschmann-2013}, to disrupting plasma confinement in tokamaks \citep{Furth-1963}. 

In addition to the importance of magnetic reconnection in driving eruptive phenomena, it is  linked to other nonlinear energy conversion processes, specifically turbulence. Turbulence is known to naturally form extended current sheets \citep{Boldyrev-2005, Chen-2012, Chandran-2015, Mallet-2016, Mallet-2017-anisotropy}, which can reconnect at sufficiently small scales \citep{Mallet-2017-recon, Mallet-2017-jpp, Loureiro-2017, Loureiro-2017prl, Vech-2018, Dong-2022}. A plethora of work based on simulations and observations has explored the intricacies that link these two fundamental processes, and how one process can lead to the generation of another \citep[e.g.][]{Stawarz-2024}.  In-situ observations provide a pathway through which we can study this.


Since its launch in 2018, Parker Solar Probe \citep[PSP;][]{Fox-2016} has observed evidence for magnetic reconnection throughout the inner heliosphere, in current sheets \citep[CSs;][]{Phan-2020} as well as at edges of switchback structures \citep{Froment-2021}. However, \citet{Phan-2020} also noted a lack of reconnection signatures in highly {\alfic} structures, showing that these non-reconnecting CSs had strong velocity shears. They suggested that large velocity shears might suppress reconnection, although the observed velocity shears were generally sub-{\alfic}, whereby the difference in the tangential flow velocity is less than the difference in the {\alf} velocity across the local current sheet. This regime had not been predicted to suppress reconnection. 

\citet{Mallet-2025} analyze the collisionless tearing mode instability in the presence of shear flow, extending earlier work on the influence of shear on the resistive tearing mode by \citet{Chen-1989}. They find that as the flow becomes more {\alfic} and shear increases, the growth rate decreases, thus suppressing reconnection in highly {\alfic} streams. This theory could potentially explain the lack of magnetic reconnection observed in highly {\alfic} near-Sun solar wind \citep{Phan-2020, Eriksson-2024}. We seek to test the analytically determined growth rate for the collisionless tearing mode instability in the presence of shear flow as derived in \citet{Mallet-2025} using near-Sun observations from PSP.

We will show that in periods identified as reconnection periods by \citet{Eriksson-2024}, the shear-modified growth rate (${\ytr}$) is large relative to the maximum growth rate without shear flow (${\ymax}$) in comparison to non-reconnection periods, even those that are not highly {\alfic}. The ratio $\yrat$ represents the relative suppression of tearing mode growth rate due to the {\alfic} shear flow. It is primarily dependent on shear velocity relative to the {\alf} speed (or the {\alf} ratio) and is related to the cross helicity and residual energy. For highly {\alfic} fast and slow wind streams, we find that ${\yrat}$ is much smaller than unity, pointing to suppression of reconnection due to shear flow as a potential reason for the lack of reconnection observed in near-Sun wind by PSP and note that reconnecting CSs are very rarely observed within these streams. We closely examine the plasma parameters around times identified as reconnection events, finding that the majority of the reconnection events are embedded in slow, non-{\alfic} wind.

The fact that reconnection current sheets are primarily found within slow, non-{\alfic} wind streams has implications for the turbulent dynamics. Work by \citet{Podesta-2010} and \citet{Bowen-2018} showed the impact of residual energy and cross helicity on the scaling of the turbulent spectra. We investigate the scaling of the spectral indices during these periods to compare with scalings in the ambient wind. 


In Section~\ref{sec:data} we outline our data selection methods and time periods studied. Section~\ref{sec:recon} includes discussion of our results related to the impact of {\alfic} shear flow on parameters associated with reconnection. In Section~\ref{sec:turb} we discuss our analysis of the spectral scaling and its relation to residual energy. In Section~\ref{sec:conc}, we outline key conclusions and implications, including ideas for future work using this dataset, and identification of areas that should be further investigated.

\section{Observations} \label{sec:data}

In this study, we use high cadence near-Sun observations from PSP to study the relative growth rate (${\yrat}$) and scaling of the magnetic field spectra. Ion and electron observations are from the ion \citep[SPANi;][]{Livi-2022} and electron \citep[SPANe;][]{Whittlesey-2020} Solar Probe ANalyzers aboard the \lq{}Solar Wind Electrons, Alphas, and Protons\rq{} \citep[SWEAP;][]{Kasper-2016} suite. Electron temperatures ($T_e$) are determined through fitting methods of the SPANe electron distributions as described in \citet{Romeo-2023, Romeo2024Thesis}. Ion temperatures ($T_i$) come from the SPANi calculated moments. SPANi observations are filtered to remove time periods where the bulk of the distribution function is outside the instrument's field of view (FOV) as this leads to non-physical and inaccurate partial moments.

Magnetic field observations come from the fluxgate magnetometer on the FIELDS instrument suite \citep{Bale-2016} providing 3d vector magnetic field measurements at high cadence. We use 4-sample per second cadenced measurements for our calculations. We resample FIELDS observations to the SPANi cadence (3.5 seconds) for our growth rate calculations, and remove time periods where the bulk of the distribution is outside the SPANi field of view. We use the full-cadence magnetic field measurements for computing spectra.

\begin{figure}[!ht]
    \centering
    \includegraphics[width=\columnwidth]{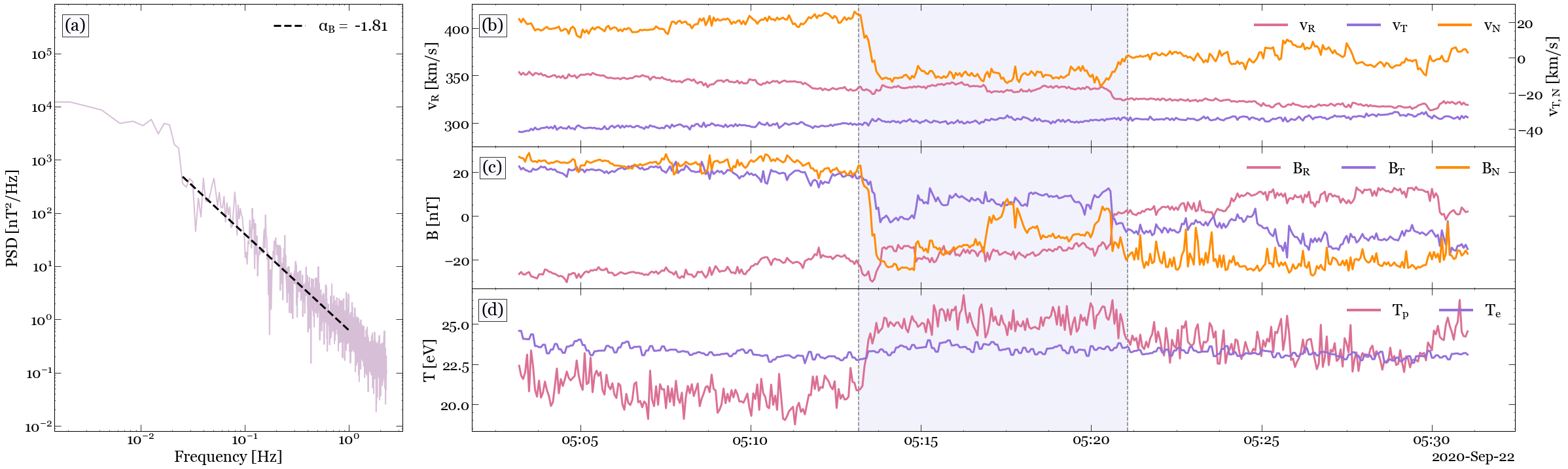}
    \caption{
    Overview of the fields and plasma parameters used in this study for the time period surrounding a reconnection event identified by \citet{Eriksson-2024} (highlighted in purple). Panel (a) shows the trace magnetic field spectrum and associated fit for the time period associated with the reconnection event (highlighted in panels (b-d)). The right side panels show the (b) SPANi proton velocity, (c) FIELDS fluxgate magnetometer magnetic field measurements, and (d) ion and electron temperatures measured by the SPANi and SPANe instruments. $v_{R, T, N}$ and $B_{R, T, N}$ are the radial, tangential, and normal components of the proton velocity (b) and magnetic field (c) respectively where R is the direction from the Sun to the spacecraft, T is the cross product of the Sun’s rotation vector with R, and N = R $\times$ T \citep{Hapgood-1992}. All data has been filtered to remove time periods where the bulk of the distribution moves out of the SPANi FOV. 
    }
    \label{fig:example_interval}
\end{figure}

We are interested in differences in the growth rate and spectral scaling between reconnection periods, and non-reconnection times. \citet{Eriksson-2024} identified reconnection periods (RPs) using a two-step process. First, they identified unique current sheets with exhaust periods that span several temporal scales. These identified periods are then checked to determine if the observed flow enhancement is consistent with a pair of propagating {\alfic} disturbances \citep{Gosling-2005}. This method led to 306 candidate periods, of which 236 has confirmed reconnection exhausts. Within each of the 236 identified periods, the SPANi FOV was checked and it was determined that the proton distribution was within the FOV of the instrument for $\geq 85 \%$ of the time in the majority of intervals \citep{Eriksson-2024}. An example reconnection period identified by \citet{Eriksson-2024} and parameters of interest are shown in Figure~\ref{fig:example_interval}. In this Figure, you can see typical signatures of reconnection such as the rotation of the field across the current sheet. In this specific event, we see an increase in the proton temperature while the electron temperature stays relatively constant (panel (d)).

\section{Suppression of Reconnection by Alfv\'enic Shear flow} \label{sec:recon}

To study the impact of {\alfic} shear flow on reconnection onset, we calculate the growth rate for the collisionless tearing mode in the presence of shear flow ($\ytr$) relative to the maximum growth rate with no shear flow ($\ymax$). This follows the analytic expression outlined in \citet{Mallet-2025}:
\begin{equation} \label{eqn:gamma}
    \frac{\gamma_{tr}}{\gamma_{0tr}} = \frac{1 - \alpha^2}{1 + \alpha^2 \tau / Z} = \frac{1 - r_A}{1 + r_A \tau / Z}  = \frac{1 - (\du / \db)^2}{1 + (\du / \db)^2 (T_i / T_e) / Z} 
\end{equation} where $\alpha = \du / \db$ is the flow shear, $\tau = T_i / T_e$ and $Z = q_i / e$. In the rest of this paper, we discuss the {\alf} ratio \citep{Chen-2013} $r_A = \alpha^2 = \du^2 / \db^2$ as to not confuse the $\alpha$ from the growth rate expression with the spectral index ($\alpha_B$) that we discuss later. In our study, we look at solely protons and electrons such that $Z = 1$ and $T_i = T_p$. $\du$ and $\db$ are the amplitudes of the fluctuations in the plasma velocity and {\alf} velocity. We calculate $\du = u - \langle u \rangle$ and $\db = b - \langle b \rangle$ where $b$ is the {\alf} velocity ($b = \frac{B}{\sqrt{\mu_0 \rho}}$). $\rho$ is calculated using a 10-minute rolling average of the proton density. $\langle \cdot \cdot \cdot \rangle$ indicates averaging of the quantity over 1 hour.

\begin{figure}[!htb]
    \centering
    \includegraphics[width=\columnwidth]{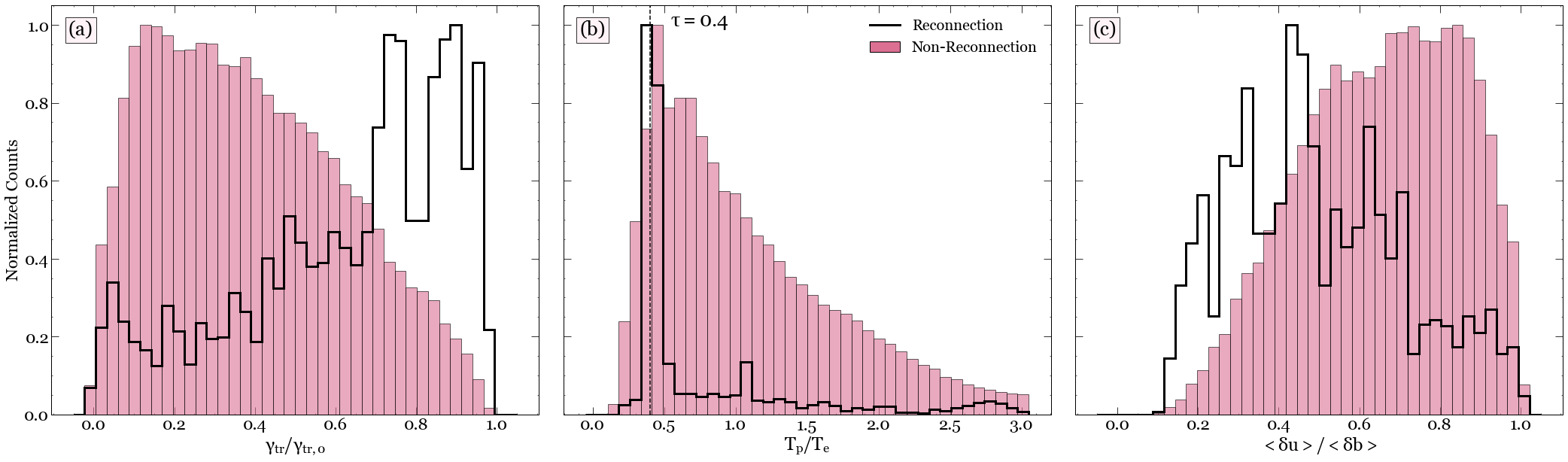}
    \caption{Comparison of (a) $\yrat$, (b) $\tau$, and (c) $\sqrt{r_A} = \alpha = \langle \du \rangle / \langle \db \rangle$ for time periods associated with reconnection (black) versus non-reconnection (pink) times. Reconnection periods are those identified by \citet{Eriksson-2024}.}
    \label{fig:recon_non}
\end{figure}

In Figure~\ref{fig:recon_non}, we look at $\yrat$, $\tau$, and $\sqrt{r_A} = \alpha$ for RPs identified by \citet{Eriksson-2024} versus non-reconnection periods (NRPs). NRPs are all background time periods within $\pm$ 5 days of perihelion for Encounters 4 through 11, where reconnection was not identified by \citet{Eriksson-2024} and the proton distribution falls within our FOV requirements \citep{Romeo-2023, Romeo2024Thesis}. 

We find that the $\yrat$ (panel (a)) for RPs to be much larger than for NRPs. This supports the idea that the inclusion of shear flow in the tearing mode growth rate expression is important for correctly understanding when the growth rate is large enough such that reconnection could occur.

In panel (b), we see that $\tau = T_p / T_e$ is tightly peaked at $\sim$ 0.4 for RPs. This could point to a preferential ion-to-electron temperature ratio that is produced through magnetic reconnection. These values are calculated within the current sheet exhaust, rather than outside the CS, and thus points to a ratio that is a by-product of the reconnection process rather than a preferential ratio for reconnection to occur. It should be noted, that we cannot directly test plasma parameters associated with reconnection onset from in-situ observations. We note that some reconnection periods show $T_p > T_e$, however the majority show higher electron temperature. \citet{Eriksson-2024} note that their identified reconnection periods are primarily embedded in wind with relatively low $T_p$, often seen in the slow solar wind. In comparison, the NRPs show a wide range of $\tau$ values, but the distribution is peaked near the $\tau$ peak for RPs. Additional examination of individual reconnection periods should be conducted in a future study. 

Panel (c) shows larger flow shear ($\alpha = \du / \db = \sqrt{r_A}$) values for NRPs in comparison to the RPs. This potentially indicates that flow shear can suppress reconnection in the ambient solar wind, and is consistent with the theoretical idea proposed by \citet{Mallet-2025}. We expand upon this to investigate the impact of different wind types on the suppression of reconnection via flow shear.

\subsection{Investigation of Relevant Plasma Parameters} \label{subsec:relative-plasma}

\begin{figure}[!htb]
    \centering
    \includegraphics[width=\columnwidth]{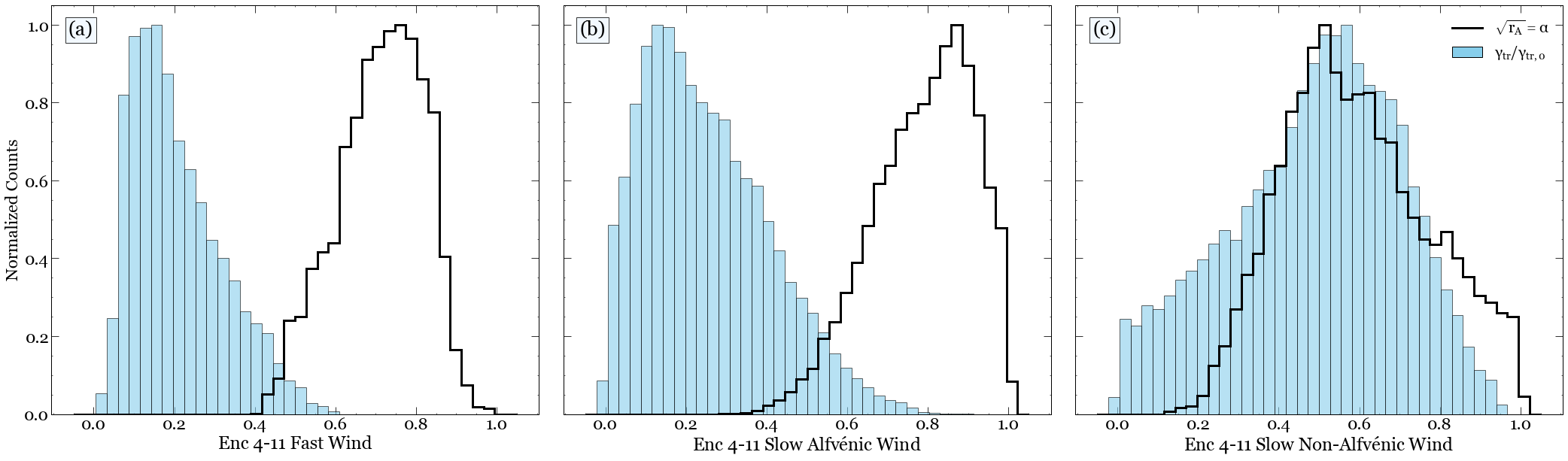}
    \caption{Comparison of $\yrat$ (blue) and normalized flow shear (black) for (a) fast wind streams, (b) slow {\alfic} wind streams, and (c) slow wind streams. Wind is identified as fast, slow {\alfic}, or slow non-{\alfic} based on the categorization scheme of \citet{Ervin-2024c}. 
    }
    \label{fig:wind_types}
\end{figure}

Figure~\ref{fig:wind_types}, compares the flow shear and $\yrat$ for fast (FSW), slow {\alfic} (SASW), and classically slow non-{\alfic} (SSW) wind as identified by \citet{Ervin-2024c}. For wind categorization, the solar wind speed and cross-helicity ($|\sigma_C|$; a proxy for {\alfty}), are used. Wind is identified as ``fast" or ``slow" using a heliocentric-distance based classification scheme outlined in \citet{Ervin-2024c}, and as ``{\alfic}" if $|\sigma_C|$ $\geq 0.7$. $\sigma_C$ is calculated following the methods of \citet{Ervin-2024b, Ervin-2024c}:

\begin{equation} \label{eqn:sigmac}
    \sigma_C = \frac{2 \langle \delta \mathbf{u} \cdot  \delta \mathbf{b} \rangle}{\langle \delta \mathbf{u}^2 \rangle + \langle \delta \mathbf{b}^2 \rangle}  = \frac{2 \sqrt{r_A}}{r_A + 1} \mathrm{cos}(\theta_{ub}) .
\end{equation}
Pure {\alf} waves correspond to $\sigma_C = \pm 1$, such that the turbulence is considered imbalanced. We note that $\sigma_C$ is dependent upon $r_A$ and thus is related to ${\yrat}$. We expect that highly {\alfic} wind should have a suppressed relative growth rate.




We find that $\yrat$ for these ambient wind streams is much lower than for the RPs (Figure~\ref{fig:recon_non}(a)). In Figure~\ref{fig:wind_types}, we show that FSW and SASW (panels (a) and (b)) have large flow shear and, in consequence per Equation~\ref{eqn:gamma}, small $\yrat$. According to the linear theory, this may explain the lack of reconnection observed by PSP in near-Sun {\alfic} wind streams. In comparison, the non-{\alfic} SSW (panel (c)) shows higher growth rates and lower shear than in the FSW and SASW. This points to reconnection periods as more likely to be found in low-{\alfty}, classically slow solar wind streams. We note that the fast wind  shows much larger temperature ratios ($\tau \geq 2$) than the slow winds ($\tau \leq 1.5$), regardless of {\alfty}. This indicates that the flow shear, which as we have shown (Equations~\ref{eqn:sigmac} and ~\ref{eqn:sigmar}) is related to the cross helicity of the plasma, is the primary parameter relevant to the growth rate.

\begin{figure}[!htb]
    \centering
    \includegraphics[width=\columnwidth]{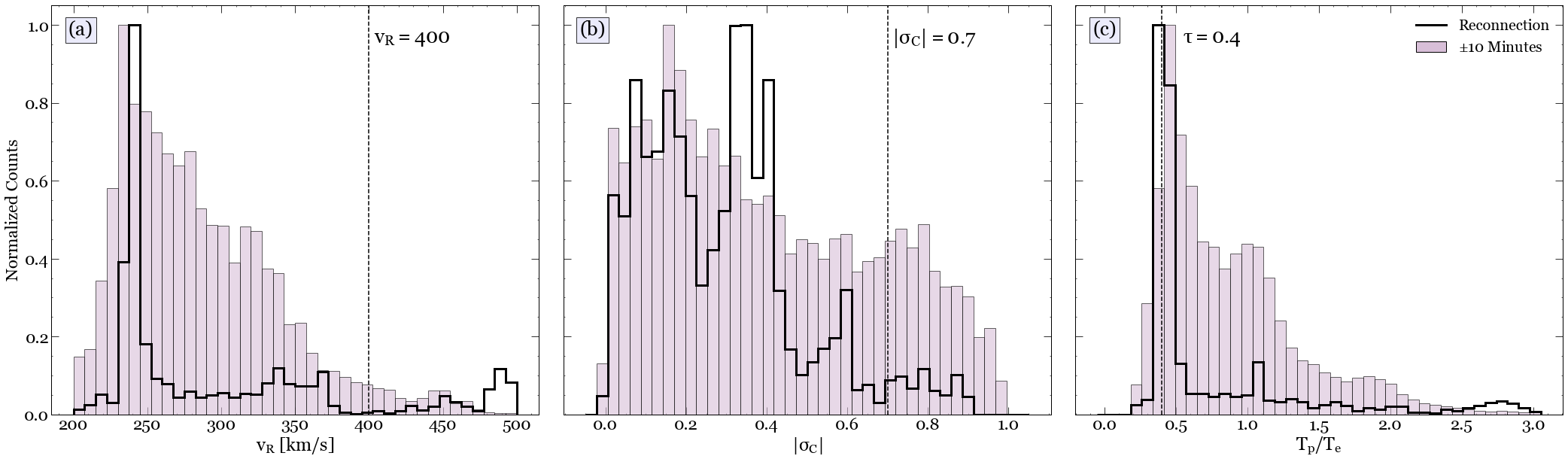}
    \caption{Comparison of (a) $v_R$, (b) $\sigma_C$, and (c) $\tau$ for time periods associated with reconnection (black) and $\pm$10-minutes near-reconnecting current sheet (NRCS) wind surrounding these events (purple).
    }
    \label{fig:surroundings}
\end{figure}

In Figure~\ref{fig:surroundings}, we look at the $\pm$10-minute time surrounding the reconnection periods identified by \citet{Eriksson-2024} to test this idea. We compare the velocity, cross helicity ($\sigma_C$), and temperature ratio ($\tau$) for wind observed $\pm$10-minutes surrounding the reconnecting periods. We refer to this $\pm$10-minute time range as ``near-reconnecting current sheet" (NRCS) periods. We note that the NRCS wind exclude the reconnection periods identified by \citet{Eriksson-2024}.

In panel (a), we find that $\sim 96 \%$ of the NRCS wind is slow (speeds less than 400 {\kms}). We see a strong peak at $\sim 245${\kms} within the reconnecting periods that could warrant further investigation. Panel (b) shows that 88\% of the NRCS wind has low {\alfty} ($|\sigma_C| \leq 0.7$). In combination, we find that the majority (85\%) of the 236 reconnecting events that were studied are embedded within non-{\alfic} SSW streams, such that $\langle|\sigma_C|\rangle \leq 0.7$ and $\langle v_R \rangle \leq 400$ {\kms} where $\langle \cdot \cdot \cdot \rangle$ is the average quantity over the NRCS interval. 

We show the ion-to-electron temperature ratio, $\tau$, in panel (c), noting that the black reconnection histogram is equivalent to that shown in Figure~\ref{fig:recon_non}. As in Figure~\ref{fig:recon_non}, we note that there is a sharp peak at $\sim 0.4$ within reconnection periods, while in the NRCS periods, we see a strong peak at $\sim 0.45$ with a larger spread. This peak is similar to the peak seen in panel (b) of Figure~\ref{fig:recon_non} for the ambient wind. This further supports our idea that the $\tau = T_i / T_e = 0.4$ ratio is a by-product of reconnection, rather than a preferential ratio for reconnection onset (again noting that we cannot directly test onset conditions at the X-lines with this in-situ dataset). The preference for RPs to be embedded in lower proton temperature wind (typical of SSW) was also noted in \citet{Eriksson-2024}.

\section{Steepening of the Magnetic Field Spectral Scaling} \label{sec:turb}
The interplay of reconnection and turbulence is important because of the generation and disruption of small scale current sheets that can lead to dissipative heating \citep[][and others]{Matthaeus-1986, Franci-2017, Mallet-2017-jpp, Loureiro-2017, Loureiro-2017prl, Huang-2016, Cerri-2017}. Turbulence is often studied through the scaling of spectra calculated from in-situ observations of the magnetic field, velocity field, etc. While the fast, high cross helicity solar wind is known to have flatter spectral scalings \citep{Chen-2020}, the slow, low cross helicity solar wind has much more variation in its observed spectral indices \citep{Bowen-2018}. Studies like \citet{Dunn-2023} have shown the wide variety of fluctuation geometries that can be generated through turbulent processes. These may potentially be related to disruption, or other processes, perhaps associated with reconnection.


As the majority of the reconnection periods have $\sigma_C \leq 0.7$, we are interested in seeing if this holds true within reconnection exhausts near the Sun and whether we can relate the generation of these steeper scalings to disruptive processes. We study the spectral index of the magnetic field spectra using PSP/FIELDS fluxgate measurements for the reconnection periods identified by \citet{Eriksson-2024}. We look at both the trace and magnitude spectral scaling to also study the generation of compressible fluctuations.

\begin{figure}[!htb]
    \centering
    \includegraphics[width=\columnwidth]{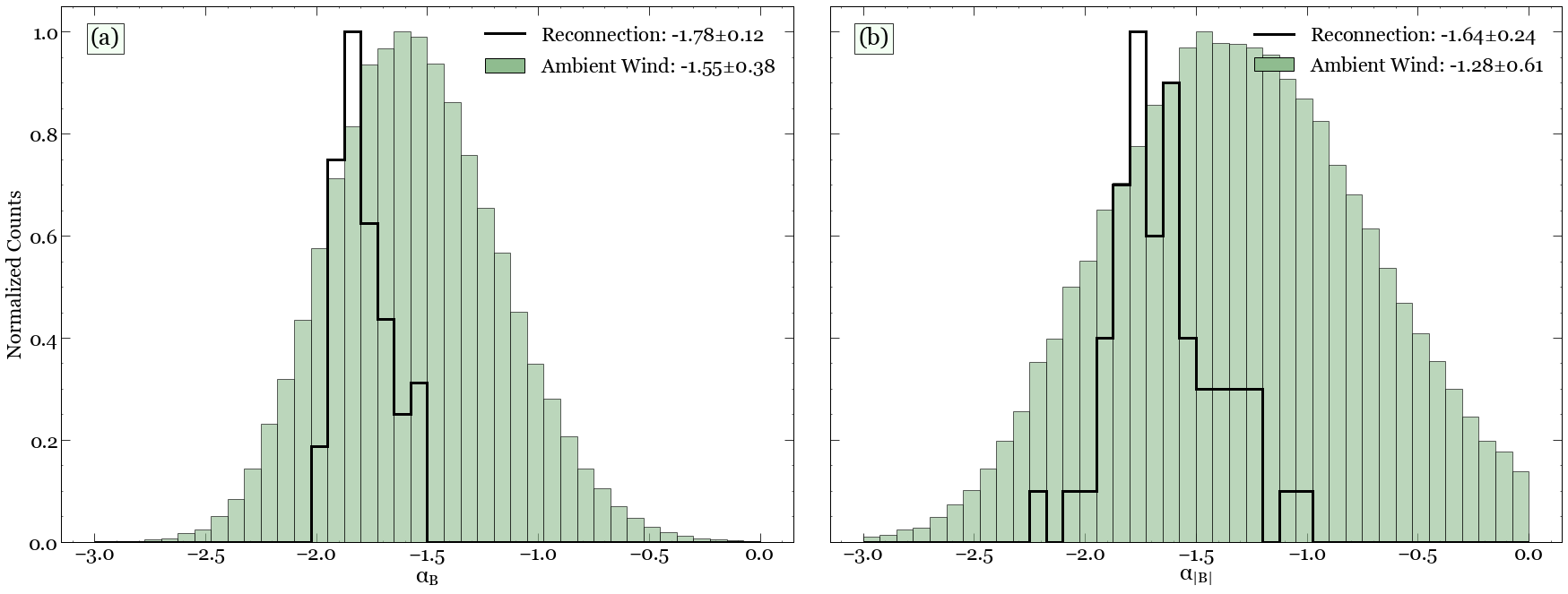}
    \caption{Comparison of magnetic field spectral index for reconnecting periods (black) identified by \citet{Eriksson-2024} and ambient wind (green). Panel (a) shows the spectral index associated with the trace spectra ($\alpha_B$) and (b) is the spectral index for the spectra of $|B|$ ($\alpha_{|B|}$). The legend reports the mean and standard deviation of spectral indices.
    }
    \label{fig:spectral_index}
\end{figure}

We calculate the power spectra for reconnection time periods longer than 1-minute, using a fast Fourier transform (FFT). Trace power spectra ($\tilde{E_B}$) are calculated as a sum of the power spectra over each of the three axes, while magnitude spectra are a FFT of $|B|$ ($\tilde{E_{|B|}}$). We do a least-square fit of the spectra and frequencies in log-log space between 0.015 and 0.25 Hz where the slope of the best fit line gives the spectral index ($\alpha$), whereby $E(f) \propto f^\alpha$. This frequency range was chosen to be far from both the outer scale and ion-scale breaks in the spectra at PSP distances, allowing us to investigate inertial range spectral scalings. We compare these indices with fits for power spectra for the ambient wind. For the ambient wind, we calculate the power spectra over 1-minute non-overlapping windows for data $\pm 5$ days around perihelion for Encounters 4 through 11 (the same time periods for NRPs discussed in Figure~\ref{fig:recon_non}). This gives us 57 reconnection spectra for comparison with 115348 spectra for ambient wind. In Figure~\ref{fig:spectral_index}, we show a comparison of the probability distribution for the fitted spectral indices and report the mean and standard deviation of the distributions. 

We find that the trace spectral indices for the reconnection periods are steeper ($\alpha_B \sim -1.78 \pm 0.12$) than for the ambient wind.  Within reconnecting periods, the plasma has low normalized residual energy ($\sigma_R \sim -1$) and the turbulence is relatively balanced ($\sigma_C \sim 0.25$). The normalized residual energy is defined as the difference in the power of the fluctuating velocity and magnetic fields: 

\begin{equation} \label{eqn:sigmar}
    \sigma_R = \frac{\langle \delta \mathbf{u}^2 \rangle - \langle \delta \mathbf{b}^2 \rangle}{\langle \delta \mathbf{u}^2 \rangle + \langle \delta \mathbf{b}^2 \rangle} =\frac{r_A - 1}{r_A + 1},
\end{equation} such that $\sigma_C^2 + \sigma_R^2 \leq 1$ geometrically. We note that, as per Eq.~\ref{eqn:sigmac}, the residual energy is directly related to the ${\yrat}$ and thus periods with smaller residual energy should show lower growth rates, and less reconnection.

The steeper spectral indices for reconnection periods are consistent with results from \citet{Bowen-2018}, where the authors noted the dependence of the magnetic field spectral index on residual energy: as the residual energy decreased, the spectra became steeper. \citet{Podesta-2010} also reported a steepening of the magnetic field spectrum with decreased $\sigma_C$, though the range of values in scaling in B can vary significantly when negative residual energy is present. 

In the ambient wind we find an average spectral index for the trace spectra of $\alpha_B \sim -1.55 \pm 0.38$. The majority of the ambient wind that PSP observes has $|\sigma_C| \geq 0.7$ and $\sigma_R \sim 0$ \citep{Ervin-2024c, Huang-2025}. These observations agree with the suggestions of \citet{Bowen-2018}, where $\alpha_B \sim -3/2$ when $\sigma_R \sim 0$.

In panel (b), we show the spectral index for the magnitude spectra ($|B|$) to look at the generation of compressible fluctuations in the solar wind. We find the spectral index within reconnection periods $\alpha_{|B|} \sim -1.64 \pm 0.24$ to be steeper than within the ambient wind $\alpha_{|B|} \sim -1.25 \pm 0.61$. A potential reason for the steepened spectra within reconnection periods is the generation of compressible fluctuations. \citet{Dunn-2023} show that the presence of compressible fluctuations leads to a steepening of the $\tilde{E}_{|B|}$ spectra to $\propto f^{-5/3}$, approximately the scaling we find within the reconnection periods. The slow wind shows large variability in plasma parameters and spectral scalings, and more work should be done to fully understand and constrain the relationship between various parameters in the slow wind and the associated spectral index.

Additional work should also be done to study the effects of reconnection on the spectral scaling of both the magnetic field and velocity spectra, and how this relates to the generation of compressible fluctuations in the solar wind. This should include the identification of additional periods to study, such that spectral scalings could be studied as a function of radial distance from the Sun \citep[e.g.][]{Sioulas-2023}. Investigation of the generation of steeper spectral scalings (e.g. -8/3 or -3) at ion scales that are theorized to be generated via reconnection processes \citep[e.g.][]{Mallet-2017-jpp} could also be done with an expanded dataset and high cadence magnetic field measurements from PSP.

\section{Conclusions} \label{sec:conc}

Through analysis of the plasma parameters and characteristics of reconnection periods using near-Sun PSP observations, we find the following:
\begin{enumerate}
    \item Times associated with reconnecting current sheets \citep[RPs;][]{Eriksson-2024} show systematically larger relative tearing-mode growth rate ($\yrat$) and smaller flow shear ($\alpha$) than for non-reconnection periods.  

    \item 85\% of reconnection periods are embedded in slow non-{\alfic} wind streams with low normalized {\alfic} flow shear (equivalently, the residual energy $\sigma_r \approx -1$). This supports the theoretical ideas put forth by \citet{Mallet-2025} that reconnection can be suppressed by {\alfic} flow shear. Highly {\alfic} slow and fast wind show large {\alfic} flow shears and small relative growth rates, with large variance in temperature ratios. This indicates that the {\alfic} flow shear is an important parameter in determining whether reconnection will occur, and  could explain the suppression of reconnection observed by PSP in the near-Sun {\alfic} solar wind by shear flow.
   
    \item Within reconnection exhausts, $\tau = T_i / T_e$ shows a small spread of values peaked at $\sim 0.4$. This potentially indicates that reconnection may favor production of this temperature ratio, as this is the ratio within the exhaust. We note that while $\tau$ appears in the relative growth rate, its precise value may depend on the specifics of the reconnection event rather than directly reflect the upstream plasma. However, $\yrat$ itself does not require such a specific value of $\tau$ and so the cause of the small spread in values is left to further investigation. This should be studied in more detail looking at individual events and their associated temperature ratios.


    \item We find that mean trace magnetic field spectral index within the reconnecting periods is larger than in the ambient wind. This is likely due to the impact of residual energy and the dynamics of balanced turbulence on the turbulent spectral index as discussed in \citet{Bowen-2018}. The magnitude spectral index within reconnection periods is found to be steeper than in the ambient wind, and consistent with scaling expected from the generation of compressive fluctuations.
    
\end{enumerate}

Our results are consistent with the idea that with strong {\alfic} flow shear, the relative growth rate of the collisionless tearing instability is small meaning the reconnection onset time would be increased. This could explain the absence of reconnection events in near-Sun {\alfic} wind observed by PSP, and indicates the importance of including flow shear in calculation of the tearing mode growth rate. Further in-depth study of the plasma parameters associated with individual events is vital to obtain a full picture of magnetic reconnection in the solar wind. 

It is important to note that this study did not calculate the absolute growth rate of the instability. In theory, one could calculate this absolute growth rate $\gamma_{tr}$ and compare to a relevant timescale of interest, for example the non-linear time ($\tau_{nl}$) or the expansion time ($\tau_{exp}$). It is unclear if these are the relevant timescales to understand the onset of reconnection (e.g. if $\tau_{nl} \gamma_{tr} \sim 1$ is the relevant parameter). Estimating the nonlinear timescale, for example, relies upon a detailed theory of the turbulence in the system. Given that we do not have a consensus theory for the imbalanced turbulence present in this system, and moreover that many assumptions about typical current sheet thicknesses and other parameters would be necessary, we have not attempted such an estimate in the present work. Future studies investigating the absolute growth rate in comparison to various timescales would be a good extension of this study to determine what is the relevant timescale, and how {\alfic} shear flow impacts reconnection onset.

In balanced turbulence, where the flux of {\alf} waves parallel and anti-parallel to the magnetic field is comparable ($\sigma_C \sim 0$), a range of $\du / \db$ values exist, with the peak of the $\du / \db$ distribution at significantly less than one, similar to the reconnecting periods in this paper (see Figure~\ref{fig:recon_non}c). Thus, the suppression of reconnection by shear is not highly effective. In contrast, in imbalanced turbulence close to the Sun, where the flux of {\alf} waves is much greater in the anti-sunward direction than the sunward direction, $\du / \db$ is instead peaked rather close to one, similar to the non-reconnecting periods in this paper (see Figure~\ref{fig:recon_non}c), amounting to a significant suppression of the reconnection rate. Thus, the prevalence of reconnection may depend crucially on the large-scale driving of the turbulence (e.g. whether it is balanced or imbalanced).

The fact that reconnection is suppressed in  environments with strong {\alfic} shear may have broader astrophysical implications. While turbulence in the interstellar medium is likely balanced (and thus $\alpha = \du / \db \sim 0$, with little suppression of reconnection at turbulence-generated CS), other astrophysical environments, for example, accretion disk coronae and outflows \citep{Chandran-2018} may be heated and accelerated by imbalanced, reflection-driven turbulence, similar to the turbulent plasma observed by PSP. Based on the evidence presented here, reconnection in turbulence-generated CS may therefore be suppressed in these situations, and this may affect the character of the heating in this turbulence, with a different partitioning of energy between ions and electrons \citep{Howes-2024}. Current sheets in the presence of flow shear also exist at planetary magnetospheric boundaries \citep{LaBelle-1994, Eriksson-2016, Sawyer-2019, Desroche-2012, Desroche-2013, DiBraccio-2013} or downstream of quasi-parallel shocks \citep{Retino-2007, Phan-2018} and our results may be of use in understanding the occurrence or suppression of reconnection in these environments as well.

\section{Acknowledgments} \label{sec: acknowledgements}
TE acknowledges funding from The Chuck Lorre Family Foundation Big Bang Theory Graduate Fellowship and NASA grant 80NSSC20K1285.
AM, SE, and MS acknowledge support from NASA grant 80NSSC20K1284. J.Juno was supported by the U.S. Department of Energy under Contract No. DE-AC02-09CH1146 via LDRD grants.
TP acknowledges support from NASA grant 80NSSC20K1781. 
TAB acknowledges support from NASA Grant 80NSSC24K0272 through the HSR program. 

The FIELDS and SWEAP experiments on the Parker spacecraft were developed and are operated under NASA contract NNN06AA01C. We acknowledge the NASA Parker Solar Probe Mission, the FIELDS team led by S. D. Bale, and the SWEAP team led by M. Stevens for use of data. 

This research used version 4.1.6 of the SunPy open source software package \citep{sunpy}, and made use of HelioPy, a community-developed Python package for space physics \citep{heliopy}.

\software{
\texttt{Astropy} \citep{astropy:2013, astropy:2018, astropy:2022},
\texttt{heliopy} \citep{heliopy},
\texttt{matplotlib} \citep{mpl},
\texttt{numpy} \citep{numpy},
\texttt{pandas} \citep{pandas},
\texttt{pySPEDAS} \citep{SPEDAS},
\texttt{scipy} \citep{scipy},
\texttt{spiceypy} \citep{spiceypy},
}

\bibliography{ms}{}

\begin{thebibliography}{}
\expandafter\ifx\csname natexlab\endcsname\relax\def\natexlab#1{#1}\fi
\providecommand{\url}[1]{\href{#1}{#1}}
\providecommand{\dodoi}[1]{doi:~\href{http://doi.org/#1}{\nolinkurl{#1}}}
\providecommand{\doeprint}[1]{\href{http://ascl.net/#1}{\nolinkurl{http://ascl.net/#1}}}
\providecommand{\doarXiv}[1]{\href{https://arxiv.org/abs/#1}{\nolinkurl{https://arxiv.org/abs/#1}}}

\bibitem[{V. Angelopoulos {et~al.}(2019)Angelopoulos, Cruce, Drozdov, Grimes, Hatzigeorgiu, King, Larson, Lewis, McTiernan, Roberts, Russell, Hori, Kasahara, Kumamoto, Matsuoka, Miyashita, Miyoshi, Shinohara, Teramoto, Faden, Halford, McCarthy, Millan, Sample, Smith, Woodger, Masson, Narock, Asamura, Chang, Chiang, Kazama, Keika, Matsuda, Segawa, Seki, Shoji, Tam, Umemura, Wang, Wang, Redmon, Rodriguez, Singer, Vandegriff, Abe, Nose, Shinbori, Tanaka, UeNo, Andersson, Dunn, Fowler, Halekas, Hara, Harada, Lee, Lillis, Mitchell, Argall, Bromund, Burch, Cohen, Galloy, Giles, Jaynes, Le~Contel, Oka, Phan, Walsh, Westlake, Wilder, Bale, Livi, Pulupa, Whittlesey, DeWolfe, Harter, Lucas, Auster, Bonnell, Cully, Donovan, Ergun, Frey, Jackel, Keiling, Korth, McFadden, Nishimura, Plaschke, Robert, Turner, Weygand, Candey, Johnson, Kovalick, Liu, McGuire, Breneman, Kersten, \& Schroeder}]{SPEDAS}
Angelopoulos, V., Cruce, P., Drozdov, A., {et~al.} 2019, \bibinfo{title}{The Space Physics Environment Data Analysis System (SPEDAS),} Space Science Reviews, 215, 9, \dodoi{10.1007/s11214-018-0576-4}

\bibitem[{A. {Annex} {et~al.}(2020){Annex}, {Pearson}, {Seignovert}, {Carcich}, {Eichhorn}, {Mapel}, {von Forstner}, {McAuliffe}, {del Rio}, {Berry}, {Aye}, {Stefko}, {de Val-Borro}, {Kulumani}, \& {Murakami}}]{spiceypy}
{Annex}, A., {Pearson}, B., {Seignovert}, B., {et~al.} 2020, \bibinfo{title}{{SpiceyPy: a Pythonic Wrapper for the SPICE Toolkit},} The Journal of Open Source Software, 5, 2050, \dodoi{10.21105/joss.02050}

\bibitem[{ {Astropy Collaboration} {et~al.}(2013){Astropy Collaboration}, {Robitaille}, {Tollerud}, {Greenfield}, {Droettboom}, {Bray}, {Aldcroft}, {Davis}, {Ginsburg}, {Price-Whelan}, {Kerzendorf}, {Conley}, {Crighton}, {Barbary}, {Muna}, {Ferguson}, {Grollier}, {Parikh}, {Nair}, {Unther}, {Deil}, {Woillez}, {Conseil}, {Kramer}, {Turner}, {Singer}, {Fox}, {Weaver}, {Zabalza}, {Edwards}, {Azalee Bostroem}, {Burke}, {Casey}, {Crawford}, {Dencheva}, {Ely}, {Jenness}, {Labrie}, {Lim}, {Pierfederici}, {Pontzen}, {Ptak}, {Refsdal}, {Servillat}, \& {Streicher}}]{astropy:2013}
{Astropy Collaboration}, {Robitaille}, T.~P., {Tollerud}, E.~J., {et~al.} 2013, \bibinfo{title}{{Astropy: A community Python package for astronomy},} \aap, 558, A33, \dodoi{10.1051/0004-6361/201322068}

\bibitem[{ {Astropy Collaboration} {et~al.}(2018){Astropy Collaboration}, {Price-Whelan}, {Sip{\H{o}}cz}, {G{\"u}nther}, {Lim}, {Crawford}, {Conseil}, {Shupe}, {Craig}, {Dencheva}, {Ginsburg}, {VanderPlas}, {Bradley}, {P{\'e}rez-Su{\'a}rez}, {de Val-Borro}, {Aldcroft}, {Cruz}, {Robitaille}, {Tollerud}, {Ardelean}, {Babej}, {Bach}, {Bachetti}, {Bakanov}, {Bamford}, {Barentsen}, {Barmby}, {Baumbach}, {Berry}, {Biscani}, {Boquien}, {Bostroem}, {Bouma}, {Brammer}, {Bray}, {Breytenbach}, {Buddelmeijer}, {Burke}, {Calderone}, {Cano Rodr{\'\i}guez}, {Cara}, {Cardoso}, {Cheedella}, {Copin}, {Corrales}, {Crichton}, {D'Avella}, {Deil}, {Depagne}, {Dietrich}, {Donath}, {Droettboom}, {Earl}, {Erben}, {Fabbro}, {Ferreira}, {Finethy}, {Fox}, {Garrison}, {Gibbons}, {Goldstein}, {Gommers}, {Greco}, {Greenfield}, {Groener}, {Grollier}, {Hagen}, {Hirst}, {Homeier}, {Horton}, {Hosseinzadeh}, {Hu}, {Hunkeler}, {Ivezi{\'c}}, {Jain}, {Jenness}, {Kanarek}, {Kendrew}, {Kern}, {Kerzendorf}, {Khvalko}, {King}, {Kirkby}, {Kulkarni},
  {Kumar}, {Lee}, {Lenz}, {Littlefair}, {Ma}, {Macleod}, {Mastropietro}, {McCully}, {Montagnac}, {Morris}, {Mueller}, {Mumford}, {Muna}, {Murphy}, {Nelson}, {Nguyen}, {Ninan}, {N{\"o}the}, {Ogaz}, {Oh}, {Parejko}, {Parley}, {Pascual}, {Patil}, {Patil}, {Plunkett}, {Prochaska}, {Rastogi}, {Reddy Janga}, {Sabater}, {Sakurikar}, {Seifert}, {Sherbert}, {Sherwood-Taylor}, {Shih}, {Sick}, {Silbiger}, {Singanamalla}, {Singer}, {Sladen}, {Sooley}, {Sornarajah}, {Streicher}, {Teuben}, {Thomas}, {Tremblay}, {Turner}, {Terr{\'o}n}, {van Kerkwijk}, {de la Vega}, {Watkins}, {Weaver}, {Whitmore}, {Woillez}, {Zabalza}, \& {Astropy Contributors}}]{astropy:2018}
{Astropy Collaboration}, {Price-Whelan}, A.~M., {Sip{\H{o}}cz}, B.~M., {et~al.} 2018, \bibinfo{title}{{The Astropy Project: Building an Open-science Project and Status of the v2.0 Core Package},} \aj, 156, 123, \dodoi{10.3847/1538-3881/aabc4f}

\bibitem[{ {Astropy Collaboration} {et~al.}(2022){Astropy Collaboration}, {Price-Whelan}, {Lim}, {Earl}, {Starkman}, {Bradley}, {Shupe}, {Patil}, {Corrales}, {Brasseur}, {N{\"o}the}, {Donath}, {Tollerud}, {Morris}, {Ginsburg}, {Vaher}, {Weaver}, {Tocknell}, {Jamieson}, {van Kerkwijk}, {Robitaille}, {Merry}, {Bachetti}, {G{\"u}nther}, {Aldcroft}, {Alvarado-Montes}, {Archibald}, {B{\'o}di}, {Bapat}, {Barentsen}, {Baz{\'a}n}, {Biswas}, {Boquien}, {Burke}, {Cara}, {Cara}, {Conroy}, {Conseil}, {Craig}, {Cross}, {Cruz}, {D'Eugenio}, {Dencheva}, {Devillepoix}, {Dietrich}, {Eigenbrot}, {Erben}, {Ferreira}, {Foreman-Mackey}, {Fox}, {Freij}, {Garg}, {Geda}, {Glattly}, {Gondhalekar}, {Gordon}, {Grant}, {Greenfield}, {Groener}, {Guest}, {Gurovich}, {Handberg}, {Hart}, {Hatfield-Dodds}, {Homeier}, {Hosseinzadeh}, {Jenness}, {Jones}, {Joseph}, {Kalmbach}, {Karamehmetoglu}, {Ka{\l}uszy{\'n}ski}, {Kelley}, {Kern}, {Kerzendorf}, {Koch}, {Kulumani}, {Lee}, {Ly}, {Ma}, {MacBride}, {Maljaars}, {Muna}, {Murphy}, {Norman},
  {O'Steen}, {Oman}, {Pacifici}, {Pascual}, {Pascual-Granado}, {Patil}, {Perren}, {Pickering}, {Rastogi}, {Roulston}, {Ryan}, {Rykoff}, {Sabater}, {Sakurikar}, {Salgado}, {Sanghi}, {Saunders}, {Savchenko}, {Schwardt}, {Seifert-Eckert}, {Shih}, {Jain}, {Shukla}, {Sick}, {Simpson}, {Singanamalla}, {Singer}, {Singhal}, {Sinha}, {Sip{\H{o}}cz}, {Spitler}, {Stansby}, {Streicher}, {{\v{S}}umak}, {Swinbank}, {Taranu}, {Tewary}, {Tremblay}, {de Val-Borro}, {Van Kooten}, {Vasovi{\'c}}, {Verma}, {de Miranda Cardoso}, {Williams}, {Wilson}, {Winkel}, {Wood-Vasey}, {Xue}, {Yoachim}, {Zhang}, {Zonca}, \& {Astropy Project Contributors}}]{astropy:2022}
{Astropy Collaboration}, {Price-Whelan}, A.~M., {Lim}, P.~L., {et~al.} 2022, \bibinfo{title}{{The Astropy Project: Sustaining and Growing a Community-oriented Open-source Project and the Latest Major Release (v5.0) of the Core Package},} \apj, 935, 167, \dodoi{10.3847/1538-4357/ac7c74}

\bibitem[{S.~D. {Bale} {et~al.}(2016){Bale}, {Goetz}, {Harvey}, {Turin}, {Bonnell}, {Dudok de Wit}, {Ergun}, {MacDowall}, {Pulupa}, {Andre}, {Bolton}, {Bougeret}, {Bowen}, {Burgess}, {Cattell}, {Chandran}, {Chaston}, {Chen}, {Choi}, {Connerney}, {Cranmer}, {Diaz-Aguado}, {Donakowski}, {Drake}, {Farrell}, {Fergeau}, {Fermin}, {Fischer}, {Fox}, {Glaser}, {Goldstein}, {Gordon}, {Hanson}, {Harris}, {Hayes}, {Hinze}, {Hollweg}, {Horbury}, {Howard}, {Hoxie}, {Jannet}, {Karlsson}, {Kasper}, {Kellogg}, {Kien}, {Klimchuk}, {Krasnoselskikh}, {Krucker}, {Lynch}, {Maksimovic}, {Malaspina}, {Marker}, {Martin}, {Martinez-Oliveros}, {McCauley}, {McComas}, {McDonald}, {Meyer-Vernet}, {Moncuquet}, {Monson}, {Mozer}, {Murphy}, {Odom}, {Oliverson}, {Olson}, {Parker}, {Pankow}, {Phan}, {Quataert}, {Quinn}, {Ruplin}, {Salem}, {Seitz}, {Sheppard}, {Siy}, {Stevens}, {Summers}, {Szabo}, {Timofeeva}, {Vaivads}, {Velli}, {Yehle}, {Werthimer}, \& {Wygant}}]{Bale-2016}
{Bale}, S.~D., {Goetz}, K., {Harvey}, P.~R., {et~al.} 2016, \bibinfo{title}{{The FIELDS Instrument Suite for Solar Probe Plus. Measuring the Coronal Plasma and Magnetic Field, Plasma Waves and Turbulence, and Radio Signatures of Solar Transients},} \ssr, 204, 49, \dodoi{10.1007/s11214-016-0244-5}

\bibitem[{S. {Boldyrev}(2005){Boldyrev}}]{Boldyrev-2005}
{Boldyrev}, S. 2005, \bibinfo{title}{{On the Spectrum of Magnetohydrodynamic Turbulence},} \apjl, 626, L37, \dodoi{10.1086/431649}

\bibitem[{T.~A. {Bowen} {et~al.}(2018){Bowen}, {Mallet}, {Bonnell}, \& {Bale}}]{Bowen-2018}
{Bowen}, T.~A., {Mallet}, A., {Bonnell}, J.~W., \& {Bale}, S.~D. 2018, \bibinfo{title}{{Impact of Residual Energy on Solar Wind Turbulent Spectra},} \apj, 865, 45, \dodoi{10.3847/1538-4357/aad95b}

\bibitem[{S.~S. {Cerri} \& F. {Califano}(2017){Cerri} \& {Califano}}]{Cerri-2017}
{Cerri}, S.~S., \& {Califano}, F. 2017, \bibinfo{title}{{Reconnection and small-scale fields in 2D-3V hybrid-kinetic driven turbulence simulations},} New Journal of Physics, 19, 025007, \dodoi{10.1088/1367-2630/aa5c4a}

\bibitem[{B.~D.~G. {Chandran} {et~al.}(2018){Chandran}, {Foucart}, \& {Tchekhovskoy}}]{Chandran-2018}
{Chandran}, B. D.~G., {Foucart}, F., \& {Tchekhovskoy}, A. 2018, \bibinfo{title}{{Heating of accretion-disk coronae and jets by general relativistic magnetohydrodynamic turbulence},} Journal of Plasma Physics, 84, 905840310, \dodoi{10.1017/S0022377818000387}

\bibitem[{B.~D.~G. {Chandran} {et~al.}(2015){Chandran}, {Schekochihin}, \& {Mallet}}]{Chandran-2015}
{Chandran}, B.~D.~G., {Schekochihin}, A.~A., \& {Mallet}, A. 2015, \bibinfo{title}{{Intermittency and Alignment in Strong RMHD Turbulence},} \apj, 807, 39, \dodoi{10.1088/0004-637X/807/1/39}

\bibitem[{C.~H.~K. {Chen} {et~al.}(2013){Chen}, {Bale}, {Salem}, \& {Maruca}}]{Chen-2013}
{Chen}, C.~H.~K., {Bale}, S.~D., {Salem}, C.~S., \& {Maruca}, B.~A. 2013, \bibinfo{title}{{Residual Energy Spectrum of Solar Wind Turbulence},} \apj, 770, 125, \dodoi{10.1088/0004-637X/770/2/125}

\bibitem[{C.~H.~K. {Chen} {et~al.}(2012){Chen}, {Mallet}, {Schekochihin}, {Horbury}, {Wicks}, \& {Bale}}]{Chen-2012}
{Chen}, C.~H.~K., {Mallet}, A., {Schekochihin}, A.~A., {et~al.} 2012, \bibinfo{title}{{Three-dimensional Structure of Solar Wind Turbulence},} \apj, 758, 120, \dodoi{10.1088/0004-637X/758/2/120}

\bibitem[{C.~H.~K. {Chen} {et~al.}(2020){Chen}, {Bale}, {Bonnell}, {Borovikov}, {Bowen}, {Burgess}, {Case}, {Chandran}, {de Wit}, {Goetz}, {Harvey}, {Kasper}, {Klein}, {Korreck}, {Larson}, {Livi}, {MacDowall}, {Malaspina}, {Mallet}, {McManus}, {Moncuquet}, {Pulupa}, {Stevens}, \& {Whittlesey}}]{Chen-2020}
{Chen}, C.~H.~K., {Bale}, S.~D., {Bonnell}, J.~W., {et~al.} 2020, \bibinfo{title}{{The Evolution and Role of Solar Wind Turbulence in the Inner Heliosphere},} \apjs, 246, 53, \dodoi{10.3847/1538-4365/ab60a3}

\bibitem[{X. Chen \& P. Morrison(1989)Chen \& Morrison}]{Chen-1989}
Chen, X., \& Morrison, P. 1989, \bibinfo{title}{Resistive tearing instability with equilibrium shear flow,} Phys. Fluids B, 2, 495

\bibitem[{T.~S. Community {et~al.}(2020)Community, Barnes, Bobra, Christe, Freij, Hayes, Ireland, Mumford, Perez-Suarez, Ryan, Shih, Contributors), Chanda, Glogowski, Hewett, Hughitt, Hill, Hiware, Inglis, Kirk, Konge, Mason, Maloney, Murray, Panda, Park, Pereira, Reardon, Savage, Sip{\H o}cz, Stansby, Jain, Taylor, Yadav, Rajul, Dang, \& Contributors)}]{sunpy}
Community, T.~S., Barnes, W.~T., Bobra, M.~G., {et~al.} 2020, \bibinfo{title}{The SunPy Project: Open Source Development and Status of the Version 1.0 Core Package,} The Astrophysical Journal, 890, 68, \dodoi{10.3847/1538-4357/ab4f7a}

\bibitem[{M. {Desroche} {et~al.}(2012){Desroche}, {Bagenal}, {Delamere}, \& {Erkaev}}]{Desroche-2012}
{Desroche}, M., {Bagenal}, F., {Delamere}, P.~A., \& {Erkaev}, N. 2012, \bibinfo{title}{{Conditions at the expanded Jovian magnetopause and implications for the solar wind interaction},} Journal of Geophysical Research (Space Physics), 117, A07202, \dodoi{10.1029/2012JA017621}

\bibitem[{M. {Desroche} {et~al.}(2013){Desroche}, {Bagenal}, {Delamere}, \& {Erkaev}}]{Desroche-2013}
{Desroche}, M., {Bagenal}, F., {Delamere}, P.~A., \& {Erkaev}, N. 2013, \bibinfo{title}{{Conditions at the magnetopause of Saturn and implications for the solar wind interaction},} Journal of Geophysical Research (Space Physics), 118, 3087, \dodoi{10.1002/jgra.50294}

\bibitem[{G.~A. {Dibraccio} {et~al.}(2013){Dibraccio}, {Slavin}, {Boardsen}, {Anderson}, {Korth}, {Zurbuchen}, {Raines}, {Baker}, {McNutt}, \& {Solomon}}]{DiBraccio-2013}
{Dibraccio}, G.~A., {Slavin}, J.~A., {Boardsen}, S.~A., {et~al.} 2013, \bibinfo{title}{{MESSENGER observations of magnetopause structure and dynamics at Mercury},} Journal of Geophysical Research (Space Physics), 118, 997, \dodoi{10.1002/jgra.50123}

\bibitem[{C. {Dong} {et~al.}(2022){Dong}, {Wang}, {Huang}, {Comisso}, {Sandstrom}, \& {Bhattacharjee}}]{Dong-2022}
{Dong}, C., {Wang}, L., {Huang}, Y.-M., {et~al.} 2022, \bibinfo{title}{{Reconnection-driven energy cascade in magnetohydrodynamic turbulence},} Science Advances, 8, eabn7627, \dodoi{10.1126/sciadv.abn7627}

\bibitem[{J. Dungey(1953)Dungey}]{Dungey-1953}
Dungey, J. 1953, \bibinfo{title}{LXXVI. Conditions for the occurrence of electrical discharges in astrophysical systems,} The London, Edinburgh, and Dublin Philosophical Magazine and Journal of Science, 44, 725, \dodoi{10.1080/14786440708521050}

\bibitem[{J.~W. {Dungey}(1961){Dungey}}]{Dungey-1961}
{Dungey}, J.~W. 1961, \bibinfo{title}{{Interplanetary Magnetic Field and the Auroral Zones},} \prl, 6, 47, \dodoi{10.1103/PhysRevLett.6.47}

\bibitem[{C. {Dunn} {et~al.}(2023){Dunn}, {Bowen}, {Mallet}, {Badman}, \& {Bale}}]{Dunn-2023}
{Dunn}, C., {Bowen}, T.~A., {Mallet}, A., {Badman}, S.~T., \& {Bale}, S.~D. 2023, \bibinfo{title}{{Effect of Spherical Polarization on the Magnetic Spectrum of the Solar Wind},} \apj, 958, 88, \dodoi{10.3847/1538-4357/ad03ef}

\bibitem[{S. {Eriksson} {et~al.}(2016){Eriksson}, {Lavraud}, {Wilder}, {Stawarz}, {Giles}, {Burch}, {Baumjohann}, {Ergun}, {Lindqvist}, {Magnes}, {Pollock}, {Russell}, {Saito}, {Strangeway}, {Torbert}, {Gershman}, {Khotyaintsev}, {Dorelli}, {Schwartz}, {Avanov}, {Grimes}, {Vernisse}, {Sturner}, {Phan}, {Marklund}, {Moore}, {Paterson}, \& {Goodrich}}]{Eriksson-2016}
{Eriksson}, S., {Lavraud}, B., {Wilder}, F.~D., {et~al.} 2016, \bibinfo{title}{{Magnetospheric Multiscale observations of magnetic reconnection associated with Kelvin-Helmholtz waves},} \grl, 43, 5606, \dodoi{10.1002/2016GL068783}

\bibitem[{S. {Eriksson} {et~al.}(2024){Eriksson}, {Swisdak}, {Mallet}, {Kruparova}, {Livi}, {Romeo}, {Bale}, {Kasper}, {Larson}, \& {Pulupa}}]{Eriksson-2024}
{Eriksson}, S., {Swisdak}, M., {Mallet}, A., {et~al.} 2024, \bibinfo{title}{{Parker Solar Probe Observations of Magnetic Reconnection Exhausts in Quiescent Plasmas near the Sun},} \apj, 965, 76, \dodoi{10.3847/1538-4357/ad25f0}

\bibitem[{T. {Ervin} {et~al.}(2024{\natexlab{a}}){Ervin}, {Jaffarove}, {Badman}, {Huang}, {Rivera}, \& {Bale}}]{Ervin-2024c}
{Ervin}, T., {Jaffarove}, K., {Badman}, S.~T., {et~al.} 2024{\natexlab{a}}, \bibinfo{title}{{Characteristics and Source Regions of Slow Alfv{\'e}nic Solar Wind Observed by Parker Solar Probe},} \apj, 975, 156, \dodoi{10.3847/1538-4357/ad7d00}

\bibitem[{T. {Ervin} {et~al.}(2024{\natexlab{b}}){Ervin}, {Bale}, {Badman}, {Bowen}, {Riley}, {Paulson}, {Rivera}, {Romeo}, {Sioulas}, {Larson}, {Verniero}, {Dewey}, \& {Huang}}]{Ervin-2024b}
{Ervin}, T., {Bale}, S.~D., {Badman}, S.~T., {et~al.} 2024{\natexlab{b}}, \bibinfo{title}{{Near Subsonic Solar Wind Outflow from an Active Region},} \apj, 972, 129, \dodoi{10.3847/1538-4357/ad57c4}

\bibitem[{N.~J. {Fox} {et~al.}(2016){Fox}, {Velli}, {Bale}, {Decker}, {Driesman}, {Howard}, {Kasper}, {Kinnison}, {Kusterer}, {Lario}, {Lockwood}, {McComas}, {Raouafi}, \& {Szabo}}]{Fox-2016}
{Fox}, N.~J., {Velli}, M.~C., {Bale}, S.~D., {et~al.} 2016, \bibinfo{title}{{The Solar Probe Plus Mission: Humanity's First Visit to Our Star},} \ssr, 204, 7, \dodoi{10.1007/s11214-015-0211-6}

\bibitem[{L. {Franci} {et~al.}(2017){Franci}, {Cerri}, {Califano}, {Landi}, {Papini}, {Verdini}, {Matteini}, {Jenko}, \& {Hellinger}}]{Franci-2017}
{Franci}, L., {Cerri}, S.~S., {Califano}, F., {et~al.} 2017, \bibinfo{title}{{Magnetic Reconnection as a Driver for a Sub-ion-scale Cascade in Plasma Turbulence},} \apjl, 850, L16, \dodoi{10.3847/2041-8213/aa93fb}

\bibitem[{C. {Froment} {et~al.}(2021){Froment}, {Krasnoselskikh}, {Dudok de Wit}, {Agapitov}, {Fargette}, {Lavraud}, {Larosa}, {Kretzschmar}, {Jagarlamudi}, {Velli}, {Malaspina}, {Whittlesey}, {Bale}, {Case}, {Goetz}, {Kasper}, {Korreck}, {Larson}, {MacDowall}, {Mozer}, {Pulupa}, {Revillet}, \& {Stevens}}]{Froment-2021}
{Froment}, C., {Krasnoselskikh}, V., {Dudok de Wit}, T., {et~al.} 2021, \bibinfo{title}{{Direct evidence for magnetic reconnection at the boundaries of magnetic switchbacks with Parker Solar Probe},} \aap, 650, A5, \dodoi{10.1051/0004-6361/202039806}

\bibitem[{H.~P. {Furth} {et~al.}(1963){Furth}, {Killeen}, \& {Rosenbluth}}]{Furth-1963}
{Furth}, H.~P., {Killeen}, J., \& {Rosenbluth}, M.~N. 1963, \bibinfo{title}{{Finite-Resistivity Instabilities of a Sheet Pinch},} Physics of Fluids, 6, 459, \dodoi{10.1063/1.1706761}

\bibitem[{J.~T. {Gosling} {et~al.}(2005){Gosling}, {Skoug}, {McComas}, \& {Smith}}]{Gosling-2005}
{Gosling}, J.~T., {Skoug}, R.~M., {McComas}, D.~J., \& {Smith}, C.~W. 2005, \bibinfo{title}{{Magnetic disconnection from the Sun: Observations of a reconnection exhaust in the solar wind at the heliospheric current sheet},} \grl, 32, L05105, \dodoi{10.1029/2005GL022406}

\bibitem[{M.~A. {Hapgood}(1992){Hapgood}}]{Hapgood-1992}
{Hapgood}, M.~A. 1992, \bibinfo{title}{{Space physics coordinate transformations: A user guide},} \planss, 40, 711, \dodoi{10.1016/0032-0633(92)90012-D}

\bibitem[{C.~R. Harris {et~al.}(2020)Harris, Millman, van~der Walt, Gommers, Virtanen, Cournapeau, Wieser, Taylor, Berg, Smith, Kern, Picus, Hoyer, van Kerkwijk, Brett, Haldane, del R{\'\i}o, Wiebe, Peterson, G{\'e}rard-Marchant, Sheppard, Reddy, Weckesser, Abbasi, Gohlke, \& Oliphant}]{numpy}
Harris, C.~R., Millman, K.~J., van~der Walt, S.~J., {et~al.} 2020, \bibinfo{title}{Array programming with NumPy,} Nature, 585, 357, \dodoi{10.1038/s41586-020-2649-2}

\bibitem[{M. {Hesse} \& P.~A. {Cassak}(2020){Hesse} \& {Cassak}}]{Hesse-2020}
{Hesse}, M., \& {Cassak}, P.~A. 2020, \bibinfo{title}{{Magnetic Reconnection in the Space Sciences: Past, Present, and Future},} Journal of Geophysical Research (Space Physics), 125, e25935, \dodoi{10.1029/2018JA025935}

\bibitem[{G.~G. {Howes}(2024){Howes}}]{Howes-2024}
{Howes}, G.~G. 2024, \bibinfo{title}{{The fundamental parameters of astrophysical plasma turbulence and its dissipation: non-relativistic limit},} Journal of Plasma Physics, 90, 905900504, \dodoi{10.1017/S0022377824001090}

\bibitem[{F. {Hoyle}(1949){Hoyle}}]{Hoyle-1949}
{Hoyle}, F. 1949, {Some recent researches in solar physics.}

\bibitem[{J. {Huang} {et~al.}(2025){Huang}, {Larson}, {Ervin}, {Liu}, {Ortiz}, {Martinovi{\'c}}, {Huang}, {Chasapis}, {Chu}, {Alterman}, {Huang}, {Wei}, {Verniero}, {Jian}, {Szabo}, {Romeo}, {Rahmati}, {Livi}, {Whittlesey}, {Alnussirat}, {Kasper}, {Stevens}, \& {Bale}}]{Huang-2025}
{Huang}, J., {Larson}, D.~E., {Ervin}, T., {et~al.} 2025, \bibinfo{title}{{The Temperature Anisotropy and Helium Abundance Features of Alfv{\'e}nic Slow Solar Wind Observed by Parker Solar Probe, Helios, and Wind Missions},} \apjl, 986, L28, \dodoi{10.3847/2041-8213/ade0ac}

\bibitem[{Y.-M. {Huang} \& A. {Bhattacharjee}(2016){Huang} \& {Bhattacharjee}}]{Huang-2016}
{Huang}, Y.-M., \& {Bhattacharjee}, A. 2016, \bibinfo{title}{{Turbulent Magnetohydrodynamic Reconnection Mediated by the Plasmoid Instability},} \apj, 818, 20, \dodoi{10.3847/0004-637X/818/1/20}

\bibitem[{J.~D. Hunter(2007)Hunter}]{mpl}
Hunter, J.~D. 2007, \bibinfo{title}{Matplotlib: A 2D graphics environment,} Computing in Science \& Engineering, 9, 90, \dodoi{10.1109/MCSE.2007.55}

\bibitem[{H. {Ji} {et~al.}(2023){Ji}, {Yoo}, {Fox}, {Yamada}, {Argall}, {Egedal}, {Liu}, {Wilder}, {Eriksson}, {Daughton}, {Bergstedt}, {Bose}, {Burch}, {Torbert}, {Ng}, \& {Chen}}]{Ji-2023}
{Ji}, H., {Yoo}, J., {Fox}, W., {et~al.} 2023, \bibinfo{title}{{Laboratory Study of Collisionless Magnetic Reconnection},} \ssr, 219, 76, \dodoi{10.1007/s11214-023-01024-3}

\bibitem[{X. {Jia} {et~al.}(2012){Jia}, {Hansen}, {Gombosi}, {Kivelson}, {T{\'o}th}, {DeZeeuw}, \& {Ridley}}]{Jia-2012}
{Jia}, X., {Hansen}, K.~C., {Gombosi}, T.~I., {et~al.} 2012, \bibinfo{title}{{Magnetospheric configuration and dynamics of Saturn's magnetosphere: A global MHD simulation},} Journal of Geophysical Research (Space Physics), 117, A05225, \dodoi{10.1029/2012JA017575}

\bibitem[{J.~C. Kasper {et~al.}(2016)Kasper, Abiad, Austin, Balat-Pichelin, Bale, Belcher, Berg, Bergner, Berthomier, Bookbinder, Brodu, Caldwell, Case, Chandran, Cheimets, Cirtain, Cranmer, Curtis, Daigneau, Dalton, Dasgupta, DeTomaso, Diaz-Aguado, Djordjevic, Donaskowski, Effinger, Florinski, Fox, Freeman, Gallagher, Gary, Gauron, Gates, Goldstein, Golub, Gordon, Gurnee, Guth, Halekas, Hatch, Heerikuisen, Ho, Hu, Johnson, Jordan, Korreck, Larson, Lazarus, Li, Livi, Ludlam, Maksimovic, McFadden, Marchant, Maruca, McComas, Messina, Mercer, Park, Peddie, Pogorelov, Reinhart, Richardson, Robinson, Rosen, Skoug, Slagle, Steinberg, Stevens, Szabo, Taylor, Tiu, Turin, Velli, Webb, Whittlesey, Wright, Wu, \& Zank}]{Kasper-2016}
Kasper, J.~C., Abiad, R., Austin, G., {et~al.} 2016, \bibinfo{title}{Solar Wind Electrons Alphas and Protons (SWEAP) Investigation: Design of the Solar Wind and Coronal Plasma Instrument Suite for Solar Probe Plus,} Space Science Reviews, 204, 131, \dodoi{10.1007/s11214-015-0206-3}

\bibitem[{J.~A. {Klimchuk}(2006){Klimchuk}}]{Klimchuk-2006}
{Klimchuk}, J.~A. 2006, \bibinfo{title}{{On Solving the Coronal Heating Problem},} \solphys, 234, 41, \dodoi{10.1007/s11207-006-0055-z}

\bibitem[{A.~L. {La Belle-Hamer} {et~al.}(1994){La Belle-Hamer}, {Otto}, \& {Lee}}]{LaBelle-1994}
{La Belle-Hamer}, A.~L., {Otto}, A., \& {Lee}, L.~C. 1994, \bibinfo{title}{{Magnetic reconnection in the presence of sheared plasma flow: Intermediate shock formation},} Physics of Plasmas, 1, 706, \dodoi{10.1063/1.870816}

\bibitem[{R. Livi {et~al.}(2022)Livi, Larson, Kasper, Abiad, Case, Klein, Curtis, Dalton, Stevens, Korreck, Ho, Robinson, Tiu, Whittlesey, Verniero, Halekas, McFadden, Marckwordt, Slagle, Abatcha, Rahmati, \& McManus}]{Livi-2022}
Livi, R., Larson, D.~E., Kasper, J.~C., {et~al.} 2022, \bibinfo{title}{The Solar Probe ANalyzer---Ions on the Parker Solar Probe,} The Astrophysical Journal, 938, 138, \dodoi{10.3847/1538-4357/ac93f5}

\bibitem[{N.~F. {Loureiro} \& S. {Boldyrev}(2017{\natexlab{a}}){Loureiro} \& {Boldyrev}}]{Loureiro-2017}
{Loureiro}, N.~F., \& {Boldyrev}, S. 2017{\natexlab{a}}, \bibinfo{title}{{Collisionless Reconnection in Magnetohydrodynamic and Kinetic Turbulence},} \apj, 850, 182, \dodoi{10.3847/1538-4357/aa9754}

\bibitem[{N.~F. {Loureiro} \& S. {Boldyrev}(2017{\natexlab{b}}){Loureiro} \& {Boldyrev}}]{Loureiro-2017prl}
{Loureiro}, N.~F., \& {Boldyrev}, S. 2017{\natexlab{b}}, \bibinfo{title}{{Role of Magnetic Reconnection in Magnetohydrodynamic Turbulence},} \prl, 118, 245101, \dodoi{10.1103/PhysRevLett.118.245101}

\bibitem[{A. Mallet {et~al.}(2025)Mallet, Eriksson, Swisdak, \& Juno}]{Mallet-2025}
Mallet, A., Eriksson, S., Swisdak, M., \& Juno, J. 2025, \bibinfo{title}{Suppression of the collisionless tearing mode by flow shear: implications for reconnection onset in the Alfvénic solar wind,} Journal of Plasma Physics, 91, E62, \dodoi{10.1017/S002237782500025X}

\bibitem[{A. {Mallet} \& A.~A. {Schekochihin}(2017){Mallet} \& {Schekochihin}}]{Mallet-2017-anisotropy}
{Mallet}, A., \& {Schekochihin}, A.~A. 2017, \bibinfo{title}{{A statistical model of three-dimensional anisotropy and intermittency in strong Alfv{\'e}nic turbulence},} \mnras, 466, 3918, \dodoi{10.1093/mnras/stw3251}

\bibitem[{A. {Mallet} {et~al.}(2017{\natexlab{a}}){Mallet}, {Schekochihin}, \& {Chandran}}]{Mallet-2017-recon}
{Mallet}, A., {Schekochihin}, A.~A., \& {Chandran}, B.~D.~G. 2017{\natexlab{a}}, \bibinfo{title}{{Disruption of sheet-like structures in Alfv{\'e}nic turbulence by magnetic reconnection},} \mnras, 468, 4862, \dodoi{10.1093/mnras/stx670}

\bibitem[{A. {Mallet} {et~al.}(2017{\natexlab{b}}){Mallet}, {Schekochihin}, \& {Chandran}}]{Mallet-2017-jpp}
{Mallet}, A., {Schekochihin}, A.~A., \& {Chandran}, B. D.~G. 2017{\natexlab{b}}, \bibinfo{title}{{Disruption of Alfv{\'e}nic turbulence by magnetic reconnection in a collisionless plasma},} Journal of Plasma Physics, 83, 905830609, \dodoi{10.1017/S0022377817000812}

\bibitem[{A. {Mallet} {et~al.}(2016){Mallet}, {Schekochihin}, {Chandran}, {Chen}, {Horbury}, {Wicks}, \& {Greenan}}]{Mallet-2016}
{Mallet}, A., {Schekochihin}, A.~A., {Chandran}, B.~D.~G., {et~al.} 2016, \bibinfo{title}{{Measures of three-dimensional anisotropy and intermittency in strong Alfv{\'e}nic turbulence},} \mnras, 459, 2130, \dodoi{10.1093/mnras/stw802}

\bibitem[{W.~H. {Matthaeus} \& S.~L. {Lamkin}(1986){Matthaeus} \& {Lamkin}}]{Matthaeus-1986}
{Matthaeus}, W.~H., \& {Lamkin}, S.~L. 1986, \bibinfo{title}{{Turbulent magnetic reconnection},} Physics of Fluids, 29, 2513, \dodoi{10.1063/1.866004}

\bibitem[{G. {Paschmann} {et~al.}(2013){Paschmann}, {{\O}ieroset}, \& {Phan}}]{Paschmann-2013}
{Paschmann}, G., {{\O}ieroset}, M., \& {Phan}, T. 2013, \bibinfo{title}{{In-Situ Observations of Reconnection in Space},} \ssr, 178, 385, \dodoi{10.1007/s11214-012-9957-2}

\bibitem[{T.~D. {Phan} {et~al.}(2018){Phan}, {Eastwood}, {Shay}, {Drake}, {Sonnerup}, {Fujimoto}, {Cassak}, {{\O}ieroset}, {Burch}, {Torbert}, {Rager}, {Dorelli}, {Gershman}, {Pollock}, {Pyakurel}, {Haggerty}, {Khotyaintsev}, {Lavraud}, {Saito}, {Oka}, {Ergun}, {Retino}, {Le Contel}, {Argall}, {Giles}, {Moore}, {Wilder}, {Strangeway}, {Russell}, {Lindqvist}, \& {Magnes}}]{Phan-2018}
{Phan}, T.~D., {Eastwood}, J.~P., {Shay}, M.~A., {et~al.} 2018, \bibinfo{title}{{Electron magnetic reconnection without ion coupling in Earth's turbulent magnetosheath},} \nat, 557, 202, \dodoi{10.1038/s41586-018-0091-5}

\bibitem[{T.~D. {Phan} {et~al.}(2020){Phan}, {Bale}, {Eastwood}, {Lavraud}, {Drake}, {Oieroset}, {Shay}, {Pulupa}, {Stevens}, {MacDowall}, {Case}, {Larson}, {Kasper}, {Whittlesey}, {Szabo}, {Korreck}, {Bonnell}, {de Wit}, {Goetz}, {Harvey}, {Horbury}, {Livi}, {Malaspina}, {Paulson}, {Raouafi}, \& {Velli}}]{Phan-2020}
{Phan}, T.~D., {Bale}, S.~D., {Eastwood}, J.~P., {et~al.} 2020, \bibinfo{title}{{Parker Solar Probe In Situ Observations of Magnetic Reconnection Exhausts during Encounter 1},} \apjs, 246, 34, \dodoi{10.3847/1538-4365/ab55ee}

\bibitem[{J.~J. {Podesta} \& J.~E. {Borovsky}(2010){Podesta} \& {Borovsky}}]{Podesta-2010}
{Podesta}, J.~J., \& {Borovsky}, J.~E. 2010, \bibinfo{title}{{Scale invariance of normalized cross-helicity throughout the inertial range of solar wind turbulence},} Physics of Plasmas, 17, 112905, \dodoi{10.1063/1.3505092}

\bibitem[{A. {Retin{\`o}} {et~al.}(2007){Retin{\`o}}, {Sundkvist}, {Vaivads}, {Mozer}, {Andr{\'e}}, \& {Owen}}]{Retino-2007}
{Retin{\`o}}, A., {Sundkvist}, D., {Vaivads}, A., {et~al.} 2007, \bibinfo{title}{{In situ evidence of magnetic reconnection in turbulent plasma},} Nature Physics, 3, 236, \dodoi{10.1038/nphys574}

\bibitem[{O.~M. Romeo(2024)Romeo}]{Romeo2024Thesis}
Romeo, O.~M. 2024, PhD thesis, University of California, Berkeley.
\newblock \url{https://www.proquest.com/dissertations-theses/interdisciplinary-approach-novel-situ/docview/3175813267/se-2}

\bibitem[{O.~M. {Romeo} {et~al.}(2023){Romeo}, {Braga}, {Badman}, {Larson}, {Stevens}, {Huang}, {Phan}, {Rahmati}, {Livi}, {Alnussirat}, {Whittlesey}, {Szabo}, {Klein}, {Niembro-Hernandez}, {Paulson}, {Verniero}, {Lario}, {Raouafi}, {Ervin}, {Kasper}, {Pulupa}, {Bale}, \& {Linton}}]{Romeo-2023}
{Romeo}, O.~M., {Braga}, C.~R., {Badman}, S.~T., {et~al.} 2023, \bibinfo{title}{{Near-Sun In Situ and Remote-sensing Observations of a Coronal Mass Ejection and its Effect on the Heliospheric Current Sheet},} \apj, 954, 168, \dodoi{10.3847/1538-4357/ace62e}

\bibitem[{R.~P. {Sawyer} {et~al.}(2019){Sawyer}, {Fuselier}, {Mukherjee}, \& {Petrinec}}]{Sawyer-2019}
{Sawyer}, R.~P., {Fuselier}, S.~A., {Mukherjee}, J., \& {Petrinec}, S.~M. 2019, \bibinfo{title}{{An Investigation of Flow Shear and Diamagnetic Drift Effects on Magnetic Reconnection at Saturn's Dawnside Magnetopause},} Journal of Geophysical Research (Space Physics), 124, 8457, \dodoi{10.1029/2019JA026696}

\bibitem[{N. {Sioulas} {et~al.}(2023){Sioulas}, {Huang}, {Shi}, {Velli}, {Tenerani}, {Bowen}, {Bale}, {Huang}, {Vlahos}, {Woodham}, {Horbury}, {de Wit}, {Larson}, {Kasper}, {Owen}, {Stevens}, {Case}, {Pulupa}, {Malaspina}, {Bonnell}, {Livi}, {Goetz}, {Harvey}, {MacDowall}, {Maksimovi{\'c}}, {Louarn}, \& {Fedorov}}]{Sioulas-2023}
{Sioulas}, N., {Huang}, Z., {Shi}, C., {et~al.} 2023, \bibinfo{title}{{Magnetic Field Spectral Evolution in the Inner Heliosphere},} \apjl, 943, L8, \dodoi{10.3847/2041-8213/acaeff}

\bibitem[{D. Stansby {et~al.}(2022)Stansby, Rai, Argall, JeffreyBroll, Haythornthwaite, Teunissen, Shaw, xypnox, Saha, Ireland, Lim, Badman, Mishra, Badger, dupuisIRT, \& tlml}]{heliopy}
Stansby, D., Rai, Y., Argall, M., {et~al.} 2022, \bibinfo{title}{heliopython/heliopy: HelioPy 1.0.0,}

\bibitem[{J.~E. {Stawarz} {et~al.}(2024){Stawarz}, {Mu{\~n}oz}, {Bessho}, {Bandyopadhyay}, {Nakamura}, {Eriksson}, {Graham}, {B{\"u}chner}, {Chasapis}, {Drake}, {Shay}, {Ergun}, {Hasegawa}, {Khotyaintsev}, {Swisdak}, \& {Wilder}}]{Stawarz-2024}
{Stawarz}, J.~E., {Mu{\~n}oz}, P.~A., {Bessho}, N., {et~al.} 2024, \bibinfo{title}{{The Interplay Between Collisionless Magnetic Reconnection and Turbulence},} \ssr, 220, 90, \dodoi{10.1007/s11214-024-01124-8}

\bibitem[{D. {Vech} {et~al.}(2018){Vech}, {Mallet}, {Klein}, \& {Kasper}}]{Vech-2018}
{Vech}, D., {Mallet}, A., {Klein}, K.~G., \& {Kasper}, J.~C. 2018, \bibinfo{title}{{Magnetic Reconnection May Control the Ion-scale Spectral Break of Solar Wind Turbulence},} \apjl, 855, L27, \dodoi{10.3847/2041-8213/aab351}

\bibitem[{P. Virtanen {et~al.}(2020)Virtanen, Gommers, Oliphant, Haberland, Reddy, Cournapeau, Burovski, Peterson, Weckesser, Bright, van~der Walt, Brett, Wilson, Millman, Mayorov, Nelson, Jones, Kern, Larson, Carey, Polat, Feng, Moore, VanderPlas, Laxalde, Perktold, Cimrman, Henriksen, Quintero, Harris, Archibald, Ribeiro, Pedregosa, van Mulbregt, Vijaykumar, Bardelli, Rothberg, Hilboll, Kloeckner, Scopatz, Lee, Rokem, Woods, Fulton, Masson, H{\"a}ggstr{\"o}m, Fitzgerald, Nicholson, Hagen, Pasechnik, Olivetti, Martin, Wieser, Silva, Lenders, Wilhelm, Young, Price, Ingold, Allen, Lee, Audren, Probst, Dietrich, Silterra, Webber, Slavi{\v c}, Nothman, Buchner, Kulick, Sch{\"o}nberger, de~Miranda~Cardoso, Reimer, Harrington, Rodr{\'\i}guez, Nunez-Iglesias, Kuczynski, Tritz, Thoma, Newville, K{\"u}mmerer, Bolingbroke, Tartre, Pak, Smith, Nowaczyk, Shebanov, Pavlyk, Brodtkorb, Lee, McGibbon, Feldbauer, Lewis, Tygier, Sievert, Vigna, Peterson, More, Pudlik, Oshima, Pingel, Robitaille, Spura, Jones, Cera, Leslie,
  Zito, Krauss, Upadhyay, Halchenko, V{\'a}zquez-Baeza, \& Contributors}]{scipy}
Virtanen, P., Gommers, R., Oliphant, T.~E., {et~al.} 2020, \bibinfo{title}{SciPy 1.0: fundamental algorithms for scientific computing in Python,} Nature Methods, 17, 261, \dodoi{10.1038/s41592-019-0686-2}

\bibitem[{ {W}es {M}c{K}inney(2010){W}es {M}c{K}inney}]{pandas}
{W}es {M}c{K}inney. 2010, in {P}roceedings of the 9th {P}ython in {S}cience {C}onference, ed. {S}t\'efan van~der {W}alt \& {J}arrod {M}illman, 56 -- 61, \dodoi{10.25080/Majora-92bf1922-00a}

\bibitem[{P.~L. Whittlesey {et~al.}(2020)Whittlesey, Larson, Kasper, Halekas, Abatcha, Abiad, Berthomier, Case, Chen, Curtis, Dalton, Klein, Korreck, Livi, Ludlam, Marckwordt, Rahmati, Robinson, Slagle, Stevens, Tiu, \& Verniero}]{Whittlesey-2020}
Whittlesey, P.~L., Larson, D.~E., Kasper, J.~C., {et~al.} 2020, \bibinfo{title}{The Solar Probe ANalyzers---Electrons on the Parker Solar Probe,} The Astrophysical Journal Supplement Series, 246, 74, \dodoi{10.3847/1538-4365/ab7370}

\end{thebibliography}
\bibliographystyle{aasjournalv7}

\end{document}